\documentclass[11pt]{article}
\usepackage{amsmath,amssymb,amsfonts}
\usepackage[english]{babel}
\makeatletter
\@addtoreset{equation}{section}
\makeatother

\def\nn{\nonumber} \def\bd{\begin{document}} \def\ed{\end{document}}
\def\ds{\documentstyle}
\let\bm=\bibitem
\newcommand{\be}{\begin{equation}}
\newcommand{\ee}{\end{equation}}
\newcommand{\bea}{\setlength\arraycolsep{2pt} \begin{eqnarray}}
\newcommand{\eea}{\end{eqnarray}}
\newcommand{\hoch}[1]{$\, ^{#1}$}
\def\p{\partial}

\title{\large {\bf A note on the definition of gravitational energy for
quadratic curvature gravity via topological regularization}}
\date{}

\author{Meng-Liang Wang$^{1}$,
\quad Jun-Jin Peng$^{2}$\footnote{corresponding author: pengjjph@163.com}\\ \\
\small \sl $^1$Guizhou Key Laboratory in Physics and Related Areas,\\
\small \sl Guizhou University of Finance and Economics,\\
\small Guiyang, Guizhou 550025, People's Republic of China; \\
\small \sl  $^2$School of Physics and Electronic Science,\\
\small \sl Guizhou Normal University, \\
\small Guiyang, Guizhou 550001, People's Republic of China
}

\begin{document}

\maketitle
\vspace{20pt}

\begin{center}
\textbf{Abstract}
\end{center}

Within the framework of four-dimensional quadratic curvature gravities in the
appearance of a negative cosmological constant, a definition for the gravitational
energy of solutions with anti-de Sitter (AdS) asymptotics was put forward in
Ref. \cite{GMOR}. This was achieved by adding proper topological invariant
terms to the gravity action to render the variation problem well-posed. In
the present note, we prove that the definition via the procedure of
topological regularization can be covered by the one given in
Ref. \cite{PLnovCC} in four dimensions. Motivated by this, we further
generalize the results to generic gravity theories in arbitrary even
dimensions.

\voffset=-.90pt
\vspace{40pt}

\section{Introduction}\label{one}

Recently, in the work \cite{GMOR}, the authors have put forward a novel
definition of the gravitational energy for the most general four-dimensional
quadratic curvature gravity theory with asymptotically anti-de Sitter (AdS) boundary
conditions, whose Lagrangian takes the form
\bea
\mathcal{L}_{QG}&=&\sqrt{-g}L_{QG} \, , \nn \\
L_{QG}&=&R-2\Lambda+\alpha R^{\mu\nu}R_{\mu\nu}+\beta R^2
\, . \label{LagQuC}
\eea
In the above equation, $\Lambda$ is the conventional negative cosmological
constant, while $\alpha$ and $\beta$ are coupling constant parameters. Their
main idea is to follow the procedure proposed in the case of general relativity
\cite{ToPReGR} to add a Gauss-Bonnet term (i.e. a topological invariant term
in four dimensions) with some weight factor to the gravity action, for the
sake of guaranteeing that the total action exhibits well-posed variational
principle when evaluated on the four-dimensional maximally symmetric solution,
that is, the global AdS$_4$ spacetime. Doing this yields the modified Iyer-Wald
prepotential \cite{IyWald}
$q^{\mu\nu}_{top}$ given by\footnote{There exists a typo in
$q^{\mu\nu}_{top}$ given by the work \cite{GMOR}.
The ``$+$" between the two terms should be ``$-$".}
\bea
q^{\mu\nu}_{top}&=&\frac{1}{2}\big(\nabla^\rho\xi^\sigma\big)
\Big[\delta^{\mu\nu}_{\rho\sigma}
+2\beta R\delta^{\mu\nu}_{\rho\sigma}
+4\alpha R^{[\mu}_{[\rho}\delta^{\nu]}_{\sigma]}
 \nn \\
&&+\frac{\ell^2}{4}(1+2\Lambda(\alpha+4\beta))
R^{\theta\omega}_{~~\gamma\lambda}
\delta^{\gamma\lambda\mu\nu}_{\theta\omega\rho\sigma}\Big]
\nn \\
&&-2\xi^\sigma\nabla^\rho\Big(
\beta R\delta^{\mu\nu}_{\rho\sigma}
+2\alpha R^{[\mu}_{[\rho}\delta^{\nu]}_{\sigma]}\Big)
\, . \label{PrePodef}
\eea
As a convention, here and in what follows a pair of square brackets on $m$
indices refer to anti-symmetrization over those indices with the common
factor of $(m!)^{-1}$, and the generalized Kronecker delta
$\delta^{\mu_1\cdot\cdot\cdot\mu_m}_{\nu_1\cdot\cdot\cdot\nu_m}$ is given by
$\delta^{\mu_1\cdot\cdot\cdot\mu_m}_{\nu_1\cdot\cdot\cdot\nu_m}=
m!\delta^{[\mu_1}_{[\nu_1}\cdot\cdot\cdot\delta^{\mu_m]}_{\nu_m]}$. For example,
$\delta^{\mu\nu}_{\rho\sigma}=\delta^{\mu}_{\rho}\delta^{\nu}_{\sigma}
-\delta^{\nu}_{\rho}\delta^{\mu}_{\sigma}$, and
$R^{[\mu}_{[\rho}\delta^{\nu]}_{\sigma]}=\big(R^{\mu}_{\rho}\delta^{\nu}_{\sigma}
-R^{\nu}_{\rho}\delta^{\mu}_{\sigma}-R^{\mu}_{\sigma}\delta^{\nu}_{\rho}
+R^{\nu}_{\sigma}\delta^{\mu}_{\rho}\big)/4$. Besides,
the radius $\ell$ of the $D$-dimensional AdS solution is associated with the negative
cosmological constant $\Lambda$ through
\be
\ell^2=-\frac{(D-1)(D-2)}{2\Lambda}
\, , \label{Radads}
\ee
while $\xi^\mu$ denotes a Killing vector field. On basis of the prepotential
(\ref{PrePodef}), the gravitational energy $Q_{top}$ for the four-dimensional
quadratic curvature gravities was proposed via the surface integral of
$q^{\mu\nu}_{top}$ in a subregion, which is given by a $(D-1)$-dimensional
hypersurface $\Sigma$ with the boundary $\partial\Sigma$, namely,
\be
Q_{top}=\frac{1}{8\pi}\int_{\partial\Sigma}
q^{\mu\nu}_{top} d\Sigma_{\mu\nu}
\, . \label{CCoftopin}
\ee
It has been testified \cite{GMOR} that $Q_{top}$ is consistent with the one
through the well-known covariant phase space method \cite{LeeWald,IyWald,WalZo}.
What is more, $Q_{top}$ yields the gravitational energy and angular momentum
of the solutions of black holes and gravitational waves in perfect agreement with the
ones via the so-called Abbott-Deser-Tekin (ADT) formalism \cite{AbbottD,DeserT}.

On the other hand, quite recently, in Ref. \cite{PLnovCC}, within the framework of
a general $D$-dimensional gravity theory described by the Lagrangian
\be
\mathcal{L}_{Riem}=\sqrt{-g}L_{Riem}
\big(g^{\alpha\beta},R_{\mu\nu\rho\sigma}\big)
\, , \label{CalLRiem}
\ee
a Komar-like formula for the conserved charges was defined through
\be
\mathcal{Q}_{Riem}=\frac{1}{8\pi}\int_{\partial\Sigma}
\mathcal{K}^{\mu\nu} d\Sigma_{\mu\nu}
\,  \label{CCofLRiem}
\ee
in terms of the 2-form potential $\mathcal{K}^{\mu\nu}$ (it is presented in
Eq. (\ref{CalKRdef}) below), provided that this gravity theory allows the
existence of the negative cosmological constant $\Lambda$ and possesses
aymptotically AdS boundary conditions.

The purpose of this short note is to demonstrate that the modified prepotential
$q^{\mu\nu}_{top}$ is identified with the 2-form $\mathcal{K}^{\mu\nu}$
in the context of the most general four-dimensional quadratic curvature
gravity theory. As a consequence, the gravitational energy
$Q_{top}=\mathcal{Q}_{Riem}$ in four dimensions. What is more, inspired with
the consistence of both the formulas in four dimensions, we are going to
extend the method of topological regularization to the generic gravity
theories described by the Lagrangian (\ref{CalLRiem}) in arbitrary even
dimensions.

\section{A comparison of the potential $q^{\mu\nu}_{top}$ with
$\mathcal{K}^{\mu\nu}$ and a generalization of  $q^{\mu\nu}_{top}$
in any even dimension}\label{two}

In the present section, we shall demonstrate explicitly that the prepotential
$q^{\mu\nu}_{top}$ for the four-dimensional quadratic curvature gravity
theories is completely consistent with $\mathcal{K}^{\mu\nu}$ in four
dimensions. Furthermore, $q^{\mu\nu}_{top}$ will be generalized to generic
gravity theories described by the Lagrangian (\ref{CalLRiem}) in arbitrary even
dimensions.

As a warmup, we follow the work \cite{PLnovCC} to present the concrete expression
of the 2-form potential $\mathcal{K}^{\mu\nu}$ in the formula (\ref{CCofLRiem}).
It is read of as
\bea
\mathcal{K}^{\mu\nu}&=&\mathcal{P}^{\mu\nu}_{~~\rho\sigma}
\nabla^\rho\xi^\sigma
-2\xi^\sigma\nabla^\rho\mathcal{P}^{\mu\nu}_{~~\rho\sigma}
\, , \label{CalKRdef}
\eea
with the tensor $\mathcal{P}^{\mu\nu\rho\sigma}$ defined through
\bea
\mathcal{P}^{\mu\nu}_{~~\rho\sigma}
&=&P^{\mu\nu}_{R\rho\sigma}
-\frac{k}{4(D-3)\hat{\Lambda}} R^{\alpha\beta}_{~~\gamma\lambda}
\delta^{\gamma\lambda\mu\nu}_{\alpha\beta\rho\sigma}
+\frac{k(D-4)}{2} \delta^{\mu\nu}_{\rho\sigma}
\,  \label{CalPRdef}
\eea
in terms of the tensor $P^{\mu\nu\rho\sigma}_R$ being of the form
\be
P^{\mu\nu\rho\sigma}_R
=\frac{\partial L_{Riem}}{\partial R_{\mu\nu\rho\sigma}}
\, . \label{PR4def}
\ee
In Eq. (\ref{CalPRdef}), both the constant parameters $\hat{\Lambda}$ and $k$
are respectively presented by
\be
\hat{\Lambda}=-\frac{1}{\ell^2} \, , \qquad
\mathcal{P}^{\mu\nu}_{~~\rho\sigma}\big(g_{\alpha\beta}
\rightarrow\bar{g}_{\alpha\beta}\big)=0
\, . \label{barPR4def}
\ee
Alternatively, the parameter $k$ can be solved from the equation
$P^{\mu\nu}_{R\rho\sigma}\big(g_{\alpha\beta}\rightarrow\bar{g}_{\alpha\beta}\big)
=k\delta^{\mu\nu}_{\rho\sigma}$, where the background metric
$\bar{g}_{\alpha\beta}$ is that of the maximally symmetric AdS spacetime
with the Riemann curvature tensor
$\bar{R}^{\mu\nu}_{\rho\sigma}(\bar{g}_{\alpha\beta})
=\hat{\Lambda} \delta^{\mu\nu}_{\rho\sigma}$.
In addition, with the help of Eq. (\ref{CalPRdef}), the potential
$\mathcal{K}^{\mu\nu}$ is able to be further expressed as
\bea
\mathcal{K}^{\mu\nu}&=& K^{\mu\nu}_R
-\frac{6k}{(D-3)\hat{\Lambda}}
R^{[\rho\sigma}_{\rho\sigma}\nabla^\mu\xi^{\nu]}
+k(D-4)\nabla^{[\mu}\xi^{\nu]} \, , \nn \\
K^{\mu\nu}_R&=&P^{\mu\nu}_{R\rho\sigma}
\nabla^\rho\xi^\sigma
-2\xi^\sigma\nabla^\rho P^{\mu\nu}_{R\rho\sigma}
\, , \label{CalKdef2}
\eea
in which the 2-form $K^{\mu\nu}_R$ is one half of the Noether potential obtained
via the covariant phase space method \cite{LeeWald,IyWald,WalZo}. It should be
pointed out that it has been illustrated \cite{PLnovCC} that the perturbation
of the potential $\mathcal{K}^{\mu\nu}$ on the background AdS spacetime
coincides with the result via the covariant phase space method, as well as that
through the (off-shell) ADT approach \cite{AbbottD,DeserT,KimKY}.

Now, we concentrate on the most general four-dimensional quadratic curvature
gravity with the Lagrangian (\ref{LagQuC}). According to the definition
(\ref{PR4def}), the rank-4 tensor $P^{\mu\nu\rho\sigma}_{QG}$ corresponding to
such a Lagrangian is read off as
\be
P^{\mu\nu}_{QG\rho\sigma}=\frac{1}{2}
\Big(\delta^{\mu\nu}_{\rho\sigma}
+2\beta R\delta^{\mu\nu}_{\rho\sigma}
+4\alpha R^{[\mu}_{[\rho}\delta^{\nu]}_{\sigma]}\Big)
\, . \label{PQGdef}
\ee
By substituting the metric $\bar{g}_{\mu\nu}$ of the background AdS$_4$ into
$P^{\mu\nu}_{QG\rho\sigma}$ and making use of
$\bar{R}^{\mu\nu}_{\rho\sigma}=\hat{\Lambda} \delta^{\mu\nu}_{\rho\sigma}$,
we arrive at
\be
P^{\mu\nu}_{QG\rho\sigma}\big(g_{\alpha\beta}\rightarrow\bar{g}_{\alpha\beta}\big)
=k_{QG}\delta^{\mu\nu}_{\rho\sigma}
\, , \label{PQGofAdS}
\ee
where the constant parameter $k_{QG}$ is given by
\bea
k_{QG}&=&\frac{1}{2}\big[1+2(D-1)(D\beta+\alpha)\hat{\Lambda}\big]
\nn \\
&=&\frac{1}{2}\big(1+2\Lambda(\alpha+4\beta)\big)
\, . \label{kQGdef}
\eea
It assists the general forth-rank tensor $\mathcal{P}^{\mu\nu}_{~~\rho\sigma}$ given
by Eq. (\ref{CalPRdef}) to turn into
\be
\mathcal{P}^{\mu\nu}_{QG\rho\sigma}
=P^{\mu\nu}_{QG\rho\sigma}
+\frac{\ell^2\big(1+2\Lambda(\alpha+4\beta)\big)}{8}
R^{\alpha\beta}_{~~\gamma\lambda}
\delta^{\gamma\lambda\mu\nu}_{\alpha\beta\rho\sigma}
\, . \label{CalPQG}
\ee
Correspondingly, the potential $\mathcal{K}^{\mu\nu}_{QG}$ for the
four-dimensional quadratic curvature gravity theories is expressed as
\bea
\mathcal{K}^{\mu\nu}_{QG}&=&\mathcal{P}^{\mu\nu}_{QG\rho\sigma}
\nabla^\rho\xi^\sigma
-2\xi^\sigma\nabla^\rho P^{\mu\nu}_{QG\rho\sigma} \nn \\
&=&q^{\mu\nu}_{top}
\, . \label{CalKQG}
\eea
This further implies that the formula for the gravitational energy via the
topological regularization scheme coincides with the formulation of the
conserved charges proposed in \cite{PLnovCC}. In contrast with the (off-shell)
ADT potential $q^{\mu\nu}_{ADT}$ \cite{DeserT,KimKY}, being of the form
\bea
q^{\mu\nu}_{ADT}
&=&\delta\big(P^{\mu\nu\rho\sigma}_{QG}
\nabla_\rho\xi_\sigma\big)-2\bar{\xi}^\sigma\bar{\nabla}^\rho
\delta P^{\mu\nu}_{QG\rho\sigma}
+k_{QG}h\bar{\nabla}^{[\mu}\bar{\xi}^{\nu]} \nn \\
&&-2k_{QG}\bar{\xi}^{[\mu}\delta^{\nu]\lambda}_{\rho\sigma}
\bar{\nabla}^\sigma h^\rho_{\lambda}
\, , \label{PerCalKQG}
\eea
where $h_{\mu\nu}=g_{\mu\nu}-\bar{g}_{\mu\nu}$, one observes that the perturbation of
$\mathcal{K}^{\mu\nu}_{QG}$ on the AdS$_4$ background, given by

\be
\delta\mathcal{K}^{\mu\nu}_{QG}=q^{\mu\nu}_{ADT}
+\frac{\ell^2k_{QG}}{2} \bar{\nabla}_\gamma
\Big[\delta^{\lambda\gamma\mu\nu}_{\alpha\beta\rho\sigma}
\Big(\bar{\nabla}^\alpha h^\beta_\lambda\Big)
\bar{\nabla}^\rho\bar{\xi}^\sigma\Big]
\, , \label{ComCKQG}
\ee
is equivalent with $q^{\mu\nu}_{ADT}$ since the difference between them is
just the divergence of a 3-form. Thus, as what has been shown in Ref. \cite{GMOR},
the formula (\ref{CCoftopin}) naturally produces the same conserved charges
as the ADT method does.

Next, under the guidance from the equivalence between the potentials $q^{\mu\nu}_{top}$
and $\mathcal{K}^{\mu\nu}$ in four dimensions, it is possible for us to
extend straightforwardly the procedure of topological regularization to
$(2n+2)$-dimensional gravity theories, where the integer $n\geq1$.
Specifically, we take into consideration of the conserved quantities for
the asymptotically AdS$_{(2n+2)}$ solutions within the context of the
gravities described by the Lagrangian (\ref{CalLRiem}) in the appearance
of the negative cosmological constant. By following Ref. \cite{PLnovCC}, we
introduce a Lovelock-type tensor $P^{\mu\nu\rho\sigma}_{(2n+2)}$  in
$2(n+1)$ dimensions, being of the form \cite{LoveGra}
\be
P^{\mu\nu}_{(2n+2)\rho\sigma}=
\frac{1}{4^n}R^{\alpha_1\beta_1}_{\gamma_1\lambda_1}\cdot\cdot\cdot
R^{\alpha_n\beta_n}_{\gamma_n\lambda_n}
\delta^{\gamma_1\lambda_1\cdot\cdot\cdot\gamma_n\lambda_n\mu\nu}
_{\alpha_1\beta_1\cdot\cdot\cdot\alpha_n\beta_n\rho\sigma}
\, , \label{PRnd}
\ee
which inherits the symmetries of the Riemann curvature tensor and is
divergence-free, that is,
$\nabla_\mu P^{\mu\nu\rho\sigma}_{(2n+2)}=0$. Apparently, the value of
$P^{\mu\nu}_{(2n+2)\rho\sigma}$ on the AdS background spacetime is given by
\be
\bar{P}^{\mu\nu}_{(2n+2)\rho\sigma}=P^{\mu\nu}_{(2n+2)\rho\sigma}
\big(g_{\alpha\beta}\rightarrow\bar{g}_{\alpha\beta}\big)
=\frac{(2n)!\hat{\Lambda}^n}{2^n}
\delta^{\mu\nu}_{\rho\sigma}
\, . \label{BarPRn}
\ee
In algebraic computations the perturbation of the
2-form $P^{\mu\nu\rho\sigma}_{(2n+2)}\nabla_\rho\xi_\sigma$ on the
AdS$_{(2n+2)}$ spacetimes gives rise to \cite{PLnovCC}
\bea
\frac{2^{n}}{(2n)!\hat{\Lambda}^{n}}
\delta\Big(P^{\mu\nu}_{(2n+2)\rho\sigma}\nabla^\rho\xi^\sigma\Big)
&=&\frac{1}{4(2n-1)\hat{\Lambda}}\delta\Big(R^{\alpha\beta}_{\gamma\lambda}
\delta^{\gamma\lambda\mu\nu}_{\alpha\beta\rho\sigma}\nabla^\rho\xi^\sigma\Big)\nn \\
&&-(n-1)\delta\big(\delta^{\mu\nu}_{\rho\sigma}
\nabla^\rho\xi^\sigma\big)
\, . \label{PertilPn}
\eea
As a consequence, in $2(n+1)$ dimensions, the tensor
$\mathcal{P}^{\mu\nu\rho\sigma}$ associated with the Lagrangian
(\ref{CalLRiem}) can be alternatively defined by
\be
\mathcal{P}^{\mu\nu}_{(2n+2)\rho\sigma}
=P^{\mu\nu}_{R\rho\sigma}
-\frac{2^{n}k}{(2n)!\hat{\Lambda}^n}
P^{\mu\nu}_{(2n+2)\rho\sigma}
\, . \label{CalPevenD}
\ee
One can check that the tensor $\mathcal{P}^{\mu\nu\rho\sigma}_{(2n+2)}$ vanishes on the
AdS$_{(2n+2)}$ spacetimes by making use of Eq. (\ref{BarPRn}). Furthermore,
the relevant potential $\mathcal{K}^{\mu\nu}$ in Eq. (\ref{CalKdef2})
becomes
\be
\mathcal{K}^{\mu\nu}_{(2n+2)}
=K^{\mu\nu}_R
-\frac{2^{n}k}{(2n)!\hat{\Lambda}^n}
P^{\mu\nu\rho\sigma}_{(2n+2)}\nabla_\rho\xi_\sigma
\, . \label{CalKevenD}
\ee
In the requirement that the generalized prepotential
$q^{\mu\nu}_{top(2n+2)}$ for the $2(n+1)$-dimensional Lagrangian
(\ref{CalLRiem}) is identified with $\mathcal{K}^{\mu\nu}_{(2n+2)}$, namely,
\be
q^{\mu\nu}_{top(2n+2)}=\mathcal{K}^{\mu\nu}_{(2n+2)}
\, ,
\ee
the supplemental topologically-invariant term to the $(2n+2)$-dimensional Lagrangian
could be proposed as
\be
\mathcal{E}_{(2n+2)}=\gamma_{(2n+2)}
\sqrt{-g}P^{\mu\nu}_{(2n+2)\rho\sigma}
R^{\rho\sigma}_{~~\mu\nu}
\, , \label{topint}
\ee
where the coupling constant $\gamma_{(2n+2)}$ is determined by
\be
\mathcal{P}^{\mu\nu\rho\sigma}_{(2n+2)}=
\frac{\partial L_{Riem}}{\partial R_{\mu\nu\rho\sigma}}
+\frac{\partial\big(\mathcal{E}_{(2n+2)}/\sqrt{-g}\big)}
{\partial R_{\mu\nu\rho\sigma}}
\, , \label{condgamm}
\ee
giving rise to
\be
\gamma_{(2n+2)}=-\frac{2^{n}k}{(2n)!(n+1)\hat{\Lambda}^n}
\, . \label{gamevend}
\ee
As an example, in the context of the most general four-dimensional quadratic
curvature gravity, $\gamma_{4}=\ell^2k_{QG}/2$ is just the coupling constant
$\gamma$ presented in Ref. \cite{GMOR}, and
$\mathcal{E}_{(4)}/(\gamma_{4}\sqrt{-g})$ becomes the Gauss-Bonnet term.
Indeed, by making use of
Eq. (\ref{gamevend}), one can follow the procedure in Ref. \cite{GMOR} to check
that the surface term from the variation of the Lagrangian $\mathcal{L}_{Riem}$
supplemented with the topological term $\mathcal{E}_{(2n+2)}$ vanishes on the
AdS$_{(2n+2)}$ spaces, rendering the variation of the Lagrangian well-behaved.

\section{Summary}\label{three}

We have demonstrated that the definition for the gravitational energy of
asymptotically AdS solutions via
topological regularization is consistent with the one presented in the work
\cite{PLnovCC} in the framework of four-dimensional quadratic curvature
gravity theories with a negative cosmological constant. What is more, we have
generalized the results in four dimensions to the generic gravity theories
with AdS asymptotics in arbitrary even dimensions. The Lagrangian describing
such gravities is supposed to be a functional built from the curvature tensor.

\section*{Acknowledgments}

This work was supported by the Natural Science Foundation of China under
Grant Nos. 11865006 and 11847152.



\begin{thebibliography}{100}

\bibitem{GMOR}
G. Giribet, O. Miskovic, R. Olea and D. Rivera-Betancour,
Energy in higher-derivative gravity via topological regularization,
Phys. Rev. D \textbf{98}, 044046 (2018).

\bibitem{ToPReGR}
R. Aros, M. Contreras, R. Olea, R. Troncoso and J. Zanelli,
Conserved charges for gravity with locally AdS asymptotics,
Phys. Rev. Lett. \textbf{84}, 1647 (2000).

\bibitem{PLnovCC}	
J.J. Peng and H.F. Liu,	
A new formula for conserved charges of Lovelock gravity in AdS spacetimes
and its generalization,
arXiv:1912.08013 [gr-qc].

\bibitem{LeeWald}
J. Lee and R. M. Wald,
Local symmetries and constraints,
J. Math. Phys. \textbf{31}, 725 (1990).

\bibitem{IyWald}
V. Iyer and R.M. Wald,
Some properties of the Noether charge and a proposal for dynamical
black hole entropy,
Phys. Rev. D \textbf{50}, 846 (1994).

\bibitem{WalZo}
R.M. Wald and A. Zoupas,
A general definition of `conserved quantities' in general relativity and
other theories of gravity,
Phys. Rev. D \textbf{61}, 084027 (2000).

\bibitem{AbbottD}
L. F. Abbott and S. Deser,
Stability of gravity with a cosmological constant,
Nucl. Phys. B \textbf{195} 76, (1982);
%
L. F. Abbott and S. Deser,
Charge definition in non-abelian gauge theories,
Phys. Lett. B \textbf{116} 259, (1982).

\bibitem{DeserT}
S. Deser and B. Tekin,
Gravitational energy in quadratic curvature gravities,
Phys. Rev. Lett. \textbf{89} 101101, (2002);
%
S. Deser and B. Tekin,
Energy in generic higher curvature gravity theories,
Phys. Rev. D \textbf{67}, 084009 (2003).

\bibitem{KimKY}
W. Kim, S. Kulkarni and S.H. Yi,
Quasi-local conserved charges in covariant theory of gravity,
Phys. Rev. Lett. \textbf{111}, 081101 (2013).

\bibitem{LoveGra}
D. Lovelock,
The Einstein tensor and its generalizations,
J. Math. Phys. \textbf{12}, 498 (1971).

\end{thebibliography}
\end{document}